\documentstyle[pre,aps,preprint]{revtex}

\begin{document}
\draft
\title{Quantum Chaos, Random Matrix Theory, and Statistical Mechanics in Two Dimensions - A Unified Approach}
\author{  Sudhir R. Jain$^{1}$ and Daniel Alonso$^{1,2}$ \\
$^1$ Facult\'e des Sciences and \\
Center for Nonlinear Phenomena and Complex Systems, \\
 Universit\'e Libre de Bruxelles,
Campus Plaine C.P. 231, \\ Boulevard du Triomphe, 
1050 Bruxelles, Belgium \\
$^2$ Departamento de F\'\i sica Fundamental y Experimental \\
Falcultad de F\'\i sicas, Universidad de La Laguna\\
La Laguna 38204, Canary Islands, Tenerife, Spain}
\maketitle

\begin{abstract}
We present a theory where the statistical mechanics for dilute ideal gases 
can be derived from random matrix approach. We show the connection of this 
approach with Srednicki approach which connects Berry conjecture with 
statistical mechanics. We further establish a link between Berry conjecture 
and random matrix theory, thus providing a unified edifice for quantum 
chaos, random matrix theory and statistical mechanics. In the course of arguing for these connections, we also observe sum rules associated with the outstanding counting problem in the theory of Braid groups.

\end{abstract}
\newpage

\noindent
{\bf 1. Introduction}\\

To understand the theme of the paper, we present an  overview of various 
different links that have been discovered in last few decades between 
classically chaotic systems and their quantal counterparts. Any study 
motivated to bring about this connection is what we understand here by 
"quantum chaos" \cite{mcg,srj}. An overwhelming number of numerical experiments on 
spectral statistics \cite{bgs,2} and their corresponding semiclassical 
analysis \cite{mvb,hdp-srj} suggest that the universal features observed in 
chaotic quantum systems can be modelled in terms of random matrix theory. 
Apart from energy spectra, it has been found that the conjecture \cite{3} 
where an eigenstate of a chaotic quantum system is represented as a Gaussian 
random superposition of plane waves entails results which are found in agreement 
with numerical studies \cite{spatial,2}. We believe that an important step has 
been in establishing the result that this conjecture leads to momentum distribution 
of ideal gases, thus bringing out statistical mechanics \cite{4}. However, in order to 
bring out the puzzling 
results in two-dimensional statistical mechanics, it is necessary that the 
choice of the correlations between amplitudes of the eigenstates be specified. 
Thus, in this pursuit, we are led to random matrix theory where one can 
systematically choose the ensemble. Recently, it has been shown how one can go 
from random matrix theory to statistical mechanics \cite{da-srj} - a  work that has brought together two important statistical theories which have 
been, hitherto, considered quite apart. 

However, there remain many questions and we describe what we believe is an 
important one. Although one can argue for thermalization \cite{4} and 
associate a temperature by a suitably-defined coarse graining, is there 
a way we can make this association more precise ? By the end of this paper, 
we hope to convince the reader that, in a certain sense, the way of explaining 
thermalization is consistent with the traditional approaches, and more 
importantly, with the second law of thermodynamics. We know that the concept 
of temperature associated with heat lies in  a curious and subtle 
combination of entropy and energy. If we could argue that the premise that 
leads us to the derivation of quantum thermalization from the Berry conjecture 
also allows us to show the acceptable behaviour of entropy, we would have a 
more coherent logical scheme tied with second law of thermodynamics. As will 
be seen, this also shows at which level of description we are with respect to 
the projection methods so widely used \cite{rau}.

Throughout the paper, we will be concentrating on two dimensions as that is 
the most difficult case in statistical mechanics \cite{5,6,7}. In section 2, 
we give a brief discussion of the choice of random matrix ensemble when 
time-reversal and parity are broken. This is fundamental in dealing 
successfully with the problem of momentum distribution function and virial 
coefficients in section 3.  The fact that quantum mechanics can be done on 
real field if the antiunitary symmetries are well-specified \cite{9}, and, 
the classification theorem of associative division algebra \cite{frob} leads 
to three basic ensembles in random matrix theory \cite{10}. Incorporating the 
violation of parity is an important step.  
In section 4, we unify the different streams of thought from quantum chaos, 
random matrix theory, and statistical mechanics by discussing entropy which 
is fundamental to all the three. We would like to mention that a recent 
work \cite{amm} is an interesting companion of this paper. We conclude the 
paper with a summary.
     
\noindent
{\bf 2.~ Random Matrix Ensemble in Two Dimensions}

In usual discussion of random matrix theory, the space dimensionality of the 
physical system plays no explicit role. Of course, it is misleading to be in 
that thought-frame. Due to the complications arising from the fact that we are 
working in two dimensions, we present here a comparative discussion about the 
fundamental symmetries in two and three (or greater) space dimensions which 
decisively restrict the possibilities of the random matrix ensemble. 

Denoting the time reversal operator by T, the position operator, q and the momentum 
operator satisfy
\begin{eqnarray}
TqT^{-1} &=& q, \nonumber \\
TpT^{-1} &=& -p.
\end{eqnarray}
In order to preserve the commutator between q and p, we see through 
\begin{equation}
TiT^{-1}=-i
\end{equation}
(i is the square root of -1) that T is antilinear. Moreover, since 
\begin{equation}
TT^{\dagger }=1,
\end{equation}
we say that T is antiunitary. T can always be written as a product of a unitary operator, U and a conjugation operator, K. On a state $\Psi $, on application of $T^2$, we get a constant $\lambda $ times $\Psi $. Note that 
\begin{equation}
T^2 = UKUK = UU^*
\end{equation}
which gives
\begin{eqnarray}
UU^* &=& \lambda , \nonumber \\
UU^{\dagger }&=& 1.
\end{eqnarray}
It now follows that $U = \lambda \overline{U}$, and hence $|\lambda |^2=1$. But then, since T is antilinear, $\lambda ^* = \lambda $ which entails 
\begin{equation}
T^2 = \pm 1.
\end{equation}
This then gives us, after proper introduction of angular momentum operator, the two possible - symmetric and antisymmetric states of a physical system which are consistent with even and half-odd integral spin respectively which, in turn, leads to Bose-Einstein and Fermi-Dirac distributions. On very general grounds thus, if a system respects time reversal symmetry, the Hamiltonian can be represented in terms of real or quaternion real elements depending on spin and rotational symmetry. If, however, time reversal is broken, the elements  are complex, and the canonical group that preserves the Hamiltonian is unitary. 
Therefore, in three (or greater) space dimesions, a random matrix ensemble can be chosen appropriately satisfying the invariance under an orthogonal, a unitary or a symplectic group, and no more \cite{frob}. 

It is important to note now that the fact that we have only symmetric or antisymmetric states here shows that the space dimensions must be three or greater since in two dimensions, there is an extra phase factor under an exchange of two coordinates which leads to a fractional angular momentum leading to fractional statistics. In the gauge where the two particles (in case of a discussion of two particles one can just consider the centre of mass as one particle) are free, the boundary conditions get twisted. For a free charged particle of charge Q in a magnetic flux, $\Phi $, the boundary condition is
$\Psi (\varphi ) \sim \exp [i(\mbox{integer} ~-~Q\Phi /2\pi )\varphi ]$; we see that the angular momentum becomes fractional \cite{saha,7}. This leads to a distinction between clockwise and anti-clockwise rotations, which leads us to the notion of chirality and the associated breakdown of parity. In general, in two space dimensions, parity and time reversal symmetries are broken. Any choice of a random matrix ensemble must be consistent with this.

Since time reversal is broken, it follows from the foregoing discussion that the Hamiltonian matrix of the system will be complex,  invariant under a unitary group. Breakdown of parity is new, however the answer is in the boundary condition. Thus, we are led to a chiral unitary ensemble. 

In our present context of many-body system whose all the eigenstates we cannot know exactly due to practical limitations (even if it is possible in some cases, we deal with the situation where a statistical study is the viable option), following Srednicki, we write a random pure state as a superposition of some basis states with amplitudes which are random. By the randomness of the amplitudes, we mean that they satisfy some correlation functions which we will write in the next section. The randomness in the amplitudes makes the pure states of the system also random. We have assumed that the system is isolated. 

The randomness in pure state can also be interpreted \cite{lubkin} by weighting the eigenvectors by a measure invariant under unitary transformations, $U_N$. By considering the unit complex N-sphere as a homogeneous space of $U_N$, then again as $U_N$ itself but organized into cosets, the measure is seen to be the Haar measure on $U_N$, thus unique. It is from this interpretation that we will discuss the entropy of the subsystem where we will note a connection between randomness in pure state and random matrix theory, however in that case it will be applied to the density operator of the subsystem  which resides in the isolated system with a Hilbert space of lesser dimensionality.

\noindent
{\bf 3. Momentum distribution } 

Let us consider a system of $N$ hard spheres ('discs' in two dimensions), 
each of radius $a$, enclosed in a box of edge-length $L+2a$. Centres of two hard 
spheres $\vec x_i$ and $\vec x_j$ are such that $|\vec x_i-\vec x_j |\ge 2 a$. 
The canonical pair of coordinates describing these particles are $(\vec X, \vec P)$ where 
$\vec X= (\vec x_1,\vec x_2, \cdots, \vec x_N), \, 
\vec P =(\vec p_1,\vec p_2, \cdots, \vec p_N)$. Energy eigenfunctions, 
$\psi_{\alpha}(\vec X)$ corresponding to eigenvalue $E_{\alpha}$ vanish on the 
boundary of the enclosure. A typical eigenfunction is irregular, with a Gaussian amplitude 
distribution and the spatial correlation function of the same is consistent with the 
conjecture of Berry which allows us to represent this eigenfunction as a superposition, following Srednicki \cite{4}: 
\begin{equation}
\label{F1}
\psi_{\alpha}(\vec X)= N_{\alpha} \int d^{dN}\vec P A_{\alpha}(\vec P) 
\delta (P^2-2mE_{\alpha})\, e^{\frac{i}{\hbar}\vec X \cdot \vec P}
\end{equation}
\noindent
with $N_{\alpha}$ given by the normalization constant, and $A_{\alpha}'s$ satisfying 
the two-point correlation function
\begin{equation}
\label{F2}
\big< A_{\alpha}^*(\vec P) A_{\gamma}(\vec P') \big>_{ME} = \delta_{\alpha \gamma}
\frac{\delta^{dN}(\vec P - \vec P')}{\delta (\vec P^2-\vec P'^2)}, 
\end{equation}
\noindent
$d$ denotes the number of coordinate-space dimensions. The average in (\ref{F2}) 
is a matrix-ensemble (ME) average which originates from the fact that the hamiltonian, 
$H$ of the system belongs to an ensemble of matrices satisfying associative  division 
algebra \cite{10,frob} in consistency with quantum mechanics. The eigenstate ensemble (EE) used in \cite{4} 
is nothing but a consequence of underlying matrix ensemble in RMT, the eigenfunctions 
then satisfy all the properties numerically observed and analytically represented 
in (\ref{F1}), (\ref{F2}) \cite{12}. The correlation functions (\ref{F2}) decide whether 
time-reversal symmetry is preserved $(A_{\alpha}^*(\vec P)=A_{\alpha}(-\vec P))$ or 
broken $(A_{\alpha}^*(\vec P) \neq A_{\alpha}(-\vec P))$, accordingly the corresponding 
matrix ensemble belongs to OE or UE respectively. As noted in \cite{4}, the higher-order 
even-point correlation functions factorize and the odd-ones vanish. A very important aspect of the ansatz 
(\ref{F1}), (\ref{F2}) is that the Wigner function corresponding to $\psi_{\alpha}(\vec X)$ is microcanonical, 
or, is proportional to $\delta (H-E_{\alpha})$ which, in a sense, incorporates ergodicity. 
We note here that, starting from an ansatz very similar to the one above, it is possible 
to obtain the quantum transport equation \cite{13} where it is important to 
relate a given quantum state with the admissible energy surface in phase 
space; thus the above ansatz is in conceptual agreement with the ergodic aspect of many-body system. 
Moreover, this choice fixes the Thomas-Fermi density of states naturally. It now becomes 
important to emphasize that we must restrict ourselves to dilute gas of hard-spheres 
and also assume that the size of sphere is much lesser than the thermal de Broglie 
wavelength. Thus, the ansatz establishes, in fact, a link between RMT and statistical mechanics. We now incorporate 
the case of two dimensions which otherwise presents enormous difficulties.

In two dimensions, the solutions of the Schr\"odinger equation, 
$\psi(\vec x_1,\vec x_2, \cdots, \vec x_N)$, under an exchange of two coordinates of particles 
satisfies
\begin{eqnarray}
\label{F3}
&& \psi(\vec x_1,\cdots, \vec x_i, \cdots, \vec x_j,\cdots , \vec x_N) \nonumber \\
= e^{i \pi \nu}
&& \psi(\vec x_1,\cdots, \vec x_j, \cdots, \vec x_i,\cdots , \vec x_N)
\end{eqnarray}
\noindent
where $\nu$ is arbitrary and defines statistics. For $\nu=0$ and $\nu=1$, with 
(\ref{F2}), one gets the Bose-Einstein and Fermi-Dirac distributions. This non-trivial phase 
and the resulting boundary condition arises from the fact that the effective 
configuration space, $M_{N}^2$ has a fundamental group, $\pi_1(M_{N}^2)=
B_N$ \cite{14}, the Braid group of $N$ objects which is an infinite, non-abelian 
group. $B_N$ is generated by $(N-1)$ elementary moves $\sigma_1, \cdots, \sigma_{N-1}$ 
satisfying the Artin relations,
\begin{eqnarray}
\label{F4}
\sigma_i\sigma_{i+1}\sigma_i=\sigma_{i+1}\sigma_i\sigma_{i+1} \, \, 
(i=1,2,\cdots, N-2) \nonumber \\
\sigma_j\sigma_i=\sigma_i\sigma_i, \, \, |i-j| \ge 2 
\end{eqnarray}
\noindent
the inverse of $\sigma_i$ is $\sigma_i^{-1}$, the identity is denoted by 
${\sl I}$, and the centre of $B_n$ is generated by $(\sigma_1\sigma_2 \cdots 
\sigma_{N-1})^N$. The multivaluedness of the eigenfunction originates from the phase 
change in  effecting an interchange between two coordinates $x_i^{(1)}$ and 
$x_i^{(2)}$ (superscripts refering to components) which can be expressed as
\begin{eqnarray}
\label{F5}
V=\exp(i \nu \sum_{i<j} \phi_{ij}), \nonumber  \\
\phi_{ij}=\tan^{-1}\big(\frac{x_i^{(2)}-x_j^{(2)}}{x_i^{(1)}-x_j^{(1)}} \big).
\end{eqnarray}

The description adopted by us here is referred to as the Anyon Gauge. It is important 
to realise that a set of coordinate configuration can be reached starting from some initial 
coordinates of $N$ particles in an infinite ways, each possibility manifested by an
action of an element $\beta \in B_N$.

The connection between initial and final sequences is given by (\ref{F3}), via 
the character $\chi(\beta)$ of the specific element. Thus, to every $\beta \in 
B_N$, we can associate the affected partial amplitude
 $\psi_{\alpha}(\beta:\vec x)$ \cite{15}. With one-dimensional unitary representation of the braid group, the 
rudiments of quantum mechanics allow us to write
\begin{equation}
\label{F6}
\Phi_{\alpha}(\vec X)=\sum_{\beta \in B_n} \chi(\beta) 
\psi_{\alpha}(\beta :\vec X)
\end{equation}
\noindent
where $\psi_{\alpha}(\beta :\vec X)$ is the probability amplitude associated in changing a 
configuration $\vec X$ to $(\beta :\vec X)$ - a configuartion after the action of $\beta$ on $\vec X$. 
The wavefunction $\Phi_{\alpha}(\vec X)$ is to be understood as appropriately 
normalised. The ansatz for $V \psi_{\alpha}(\beta :\vec X)$ is now
\begin{equation}
\label{F7}
V \psi_{\alpha}(\beta :\vec X)=
N_{\alpha} \int d^{2N}\vec P A_{\alpha}(\beta: \vec P) 
\delta (P^2-2mE_{\alpha}) e^{\frac{i}{\hbar}\vec X \cdot \vec P}
\end{equation}
\noindent
with $A_{\alpha}(\beta: \vec P)$ satisfying
\begin{equation}
\label{F8}
\big< A_{\alpha}^*(\beta_1:\vec P_1) A_{\gamma}(\beta_2:\vec P_2) \big>_{ME} =
 \delta_{\alpha \gamma}
\frac{\delta^{2N}\big((\beta_1:\vec P_1) - (\beta_2:\vec P_2)\big)}
{\delta (\vec P^2_1-\vec P^2_2)},
\end{equation}
\noindent
$(\beta_1,\beta_2 \in B_N)$, and $A_{\alpha}(\vec P)$ satisfy the twisted boundary
conditions,
\begin{eqnarray}
\label{F9}
&& A_{\alpha}(\vec p_1,\cdots, \vec p_i, \cdots, \vec p_j,\cdots , \vec p_N) \nonumber \\ 
 = e^{i \pi \nu}
&& A_{\alpha}(\vec p_1,\cdots, \vec p_j, \cdots, \vec p_i,\cdots , \vec p_N)
\end{eqnarray}

The question now is in specifying exactly what the 
matrix ensemble is in this case? The form of (\ref{F8}) with $A_{\alpha}$'s not 
restricted to real, takes into account the T-breaking, and (\ref{F9}) makes the 
ensemble handed or chiral as a result of P-breaking. Thus (\ref{F7})-(\ref{F9}) gives 
the complete description and the ME is, in fact, the chiral-Gaussian Unitary Ensemble 
(ch-GUE) \cite{16} as discussed in the previous section from general considerations. It can be easily shown that the Wigner distribution is
\begin{eqnarray}
\label{10}
\big<\rho_{\alpha}^W(\vec X , \vec P) \big>_{ME}=n_{\alpha}^{-1} h^{-2N} 
\delta(\frac{P^2}{2m}-E_{\alpha}), \nonumber \\
n_{\alpha}= \frac{1}{N! \Gamma(N) E_{\alpha}}\big( 
\frac{m L^2 E_{\alpha}}{2 \pi \hbar^2} \big)^N.
\end{eqnarray}

For the momentum distribution, we need to evaluate the ME-average of 
$\tilde \Phi_{\alpha}^*(\vec P)\tilde \Phi_{\gamma}^*(\vec P")$ with $\tilde \Phi 
\equiv V\Phi_{\alpha}$. With the above ansatz and conditions supplementing it, this average is
\begin{eqnarray}
\label{F11}
&& {\cal{F}}(\vec P)= 
\big< \tilde \Phi_{\alpha'}^*(\vec P)\tilde \Phi_{\gamma}^*(\vec P") \big>_{ME} = \nonumber \\
&& h^{2N}\delta_{\alpha' \gamma} N_{\alpha '} N_{\gamma} 
\sum_{n,m=0} \sum_{\beta_1(m)} \sum_{\beta_2(n)} \chi^*(\beta_1) \chi(\beta_2) 
\delta (P^2-2mE_{\alpha '}) \nonumber \\ 
&& \times \delta_{\cal D}^{2N} \bigg( \prod_{\alpha =0}^m 
\sigma_{\beta_1(\alpha )}^{\epsilon_{\beta_1}} \vec P"  -
      \prod_{\alpha =0}^n
\sigma_{\beta_2(\alpha )}^{\epsilon_{\beta_2}} \vec P 
\bigg) \bigg|_{\vec P = \vec P"} 
\end{eqnarray}
\noindent
where
\begin{equation}
\label{F12}
\delta_{{\cal D}}^{2N}(\vec Q)= h^{-2N} 
\int_{Domain, \cal D} d^{2N}X \exp \big( \frac{i}{\hbar} \vec Q \cdot \vec X \big) ;
\end{equation}
\noindent
$\vec P$ is identified with $\vec P"$ after the sum is performed.

With (\ref{F11}), the momentum distribution is given by
\begin{equation}
F(\vec p_1)= \frac{\int d \vec p_2 \cdots d \vec p_N {\cal F}(\vec P)}
                  {\int d \vec p_1 \cdots d \vec p_N {\cal F}(\vec P)}
\end{equation}
\noindent
which formally completes the deduction. However, an exact evaluation of this is 
very difficult and the difficulty is coming from  counting of irreducible 
words formed by the $\sigma 's$. To make the precise connection, we give derive the result 
upto $O(\hbar^2/L^2)$, an order that is enough for second virial coefficient. 

In deriving the momentum distribution, we have to consider all the exchanges that lead to contributions giving second virial coefficient. With N generators, we have characters $e^{i\pi \nu}$ and $e^{-i\pi \nu}$ leading to a combination,
cos($N\pi \nu$). With one of the momenta fixed in the above integral, we look for elements of $B_N$ such that two momenta are interchanged restoring all other momenta to their labellings. All these elements contribute upto $O(\hbar ^2/L^2)$. There are two kinds of terms with any $\sigma _M$ (which denotes the elements of $B_N$ with M generators):

\noindent 
(i) one where $\vec{p_1}$ changes,\\
\noindent
(ii) one where $\vec{p_1}$ does not change.\\

We first insert a notation which will be used in sequel, viz., the integral,
\begin{equation}
\int d^Np\delta (p^2 - x) := I_N(x) = \frac{(\pi x)^{N/2}}{\Gamma (N/2)x}.
\end{equation}
In case (ii), we have typically
\begin{equation}
\delta _{D}(\vec p_1 - \vec p_1)\delta _{D}(\vec p_2 - \vec p_2)...\delta _{D}(\vec p_j - \vec p_i)\delta _{D}(\vec p_i - \vec p_j)...\delta _{D}(\vec p_N - \vec p_N). 
\end{equation}
This leads to the value of the integral,
\begin{equation}
{1 \over 2}I_{2(N-2)}(2mE_{\alpha }-p_1^2)\left( \frac{L}{h}\right)^{2(N-1)}{\cal R}_{{\vec \sigma }_{M}}(N)\chi ({{\vec \sigma }_{M}})
\end{equation}
where ${\cal R}_{{\vec \sigma }_{M}} $ denotes the number of elements composed by M generators that contribute to $O(h^2/L^2)$, or just an interchange between two momenta but not $\vec p_1$, and, $\chi ({{\vec \sigma }_{M}})$ denotes the corresponding character. 

In case (i), we have typically
\begin{equation}
\delta _{D}(\vec p_2 - \vec p_1)\delta _{D}(\vec p_1 - \vec p_2)...\delta _{D}(\vec p_N - \vec p_N). 
\end{equation}
which leads to the integral evaluating to
\begin{equation}
I_{2(N-2)}(2mE_{\alpha }-2p_1^2)\left( \frac{L}{h}\right)^{2(N-1)}
{\cal Q}_{{\vec \sigma }_{M}}(N)\chi ({{\vec \sigma }_{M}})
\end{equation}
where ${\cal Q}_{{\vec \sigma }_{M}}(N)$ denotes the number of elements of $B_N$ composed by N 
generators contributing to $O(h^2/L^2)$ that involve an interchange with $\vec p_1$.

For each ${\vec \sigma }_{M} \rightarrow \chi ({\vec \sigma }_{M})$, we can find 
${\vec \sigma }_{M}^* \rightarrow \chi ^*({\vec \sigma }_{M})$, and 
$({\cal Q}_{{\vec \sigma }_{M}},{\cal R}_{{\vec \sigma }_{M}}) = 
({\cal Q}_{{\vec \sigma }_{M}^{*}},{\cal R}_{{\vec \sigma }_{M}^{*}}) $. 
Thus, for fixed M, case (i) gives
\begin{equation}
\left(\frac{L}{h}\right)^{2(N-1)}I_{2(N-2)}(2mE_{\alpha }-2p_1^2)
{\cal Q}_{{\vec \sigma }_{M}}(N)\left( \chi ({{\vec \sigma }_{M}}) + 
\chi ^{*}({{\vec \sigma }_{M}}) \right);
\end{equation}
and case (ii) gives
\begin{equation}
\left(\frac{L}{h}\right)^{2(N-1)}{1 \over 2}I_{2(N-2)}(2mE_{\alpha }-p_1^2)
    {\cal R}_{{\vec \sigma }_{M}}(N)
\big( \chi ({{\vec \sigma }_{M}}) + 
    \chi^{*}({{\vec \sigma }_{M}}) \big).
\end{equation}
Because only two momenta are interchanged, the total contribution of elements of $B_N$ 
formed by M generators is
\begin{eqnarray}
\left(\frac{L}{h}\right)^{2(N-1)}\sum_{k=-M,-M+2,...,M-2,M} 
 \Bigl[ I_{2(N-2)}(2mE_{\alpha }- p_1^2)
    \cos (\pi k\nu){\cal R}_k^{(M)}(N) \nonumber \\
+     2I_{2(N-2)}(2mE_{\alpha }-2p_1^2) 
    \cos (\pi k\nu){\cal Q}_k^{(M)}(N) \Bigr]
\end{eqnarray}
for
\begin{equation}
\chi ({\vec \sigma }_{M}) = \Bigl[ e^{-iM\pi \nu},e^{-i(M-2)\pi \nu},...,
e^{iM\pi \nu} \Bigr] = e^{ik\pi \nu},
\end{equation}
and, ${\cal R}_{{\vec \sigma }_{M}}$is just ${\cal R}_k^{(M)}(N)$.
If we integrate over $\vec p_1$, we obtain the normalization factor. For 
this, we have terms that lead to exchange as discussed above, and also the 
operation of identity of $B_N$ where no momenta are changed. To begin with, 
we have the integration of the term without identity, and the result is 
\begin{eqnarray}
\left(\frac{L}{h}\right)^{2(N-1)}\sum_{k=-M,-M+2,...,M-2,M} 
\Bigl[ I_{2(N-1)}(2mE_{\alpha })\cos (\pi k\nu){\cal R}_k^{(M)}(N)  \nonumber \\ 
+ 2I_{2(N-1)}(2mE_{\alpha })\cos (\pi k\nu){\cal Q}_k^{(M)}(N) \Bigr].
\end{eqnarray}

Denoting by ${\cal P}_{{\vec \sigma }_{M}}(N)$ by the number of elements of $B_N$ composed of M generators contributing to $O(h^0/L^0)$ - the identity, integration over $\vec p_2...\vec p_N$ gives
\begin{equation}
\left(\frac{L}{h}\right)^{2N}I_{2(N-1)}(2mE_{\alpha }-p_1^2){\cal P}_{{\vec \sigma }_{M}}(N)\chi ({\vec \sigma }_{M}).
\end{equation}
As above, we have ${\vec \sigma }_{M}$ and ${\vec \sigma }_{M}^{*}$, so this integral reduces to
\begin{equation}
\left(\frac{L}{h}\right)^{2N}\sum_{k=-M,-M+2,...,M-2,M}2I_{2(N-1)}(2mE_{\alpha }-p_1^2)\cos (\pi k\nu){\cal P}_{k}^{M}(N).
\end{equation}
To get the contribution of identity to normalization, we now integrate this over $\vec p_1$ to obtain
\begin{equation}
\left(\frac{L}{h}\right)^{2N}\sum_{k=-M,-M+2,...,M-2,M}2I_{2N}(2mE_{\alpha })\cos (\pi k\nu){\cal P}_{k}^{M}(N).
\end{equation}

For the second virial coefficient, if M = 2m (m=0,1,2,...), the contribution goes 
to the term involved in  identity, and, if M = 2m + 1 (m=0,1,2,...), the contribution 
is $O(h^2/L^2)$. The sum over elements of $B_N$ can be substituted by a sum over m. 
All put together, in the term which gives normalization, we have 
\begin{equation}
O(1) ~:~\left(\frac{L}{h} \right)^{2N}\sum_{m=0}^{\infty } 
~\sum_{k=-2m,-2m+2,...2m}2I_{2N}(2mE_{\alpha })\cos (\pi k\nu){\cal P}_{k}^{M}(N);
\end{equation}
and
\begin{eqnarray}
O\left( \frac{h^2}{L^2}\right)~:~\left(\frac{L}{h} \right)^{2(N-1)}
\sum_{m=0}^{\infty } ~\sum_{k=-2m-1,-2m+1,...2m+1}
\Bigl[ I_{2(N-1)}(2mE_{\alpha })\cos (\pi k\nu){\cal R}_{k}^{M}(N)\nonumber \\
+  I_{2(N-1)}(2mE_{\alpha })\cos (\pi k\nu){\cal Q}_{k}^{M}(N) \Bigr].
\end{eqnarray}
For the numerator of (19), with one momentum $\vec p_1$ fixed and integrating with respect to all other momenta, we get the following results :
\begin{equation}
O(1) ~:~\left(\frac{L}{h} \right)^{2N}\sum_{m=0}^{\infty } ~\sum_{k=-2m,-2m+2,...2m}2I_{2(N-1)}(2mE_{\alpha }-p_1^2)\cos (\pi k\nu){\cal P}_{k}^{M}(N);
\end{equation}
and
\begin{eqnarray}
O\left( \frac{h^2}{L^2}\right)~:~\left(\frac{L}{h} 
\right)^{2(N-1)}\sum_{m=0}^{\infty }~ \sum_{k=-2m-1,-2m+1,...2m+1}
\Bigl[ I_{2(N-2)}(2mE_{\alpha }-p_1^2)\cos (\pi k\nu){\cal R}_{k}^{M}(N)\nonumber \\
+  2I_{2(N-2)}(2mE_{\alpha }-2p_1^2)\cos (\pi k\nu)
{\cal Q}_{k}^{M}(N) \Bigr].
\end{eqnarray}

For large N, 
\begin{equation}
I_{2(N-1)}(x) ~\sim ~ I_{2N}(x)
\end{equation}
as $I_N$(x) is just the volume of an N-dimensional sphere of radius x. Let us define
\begin{eqnarray}
A &=& \frac{p_1^2}{2mk_BT_{\alpha }}, \nonumber \\
B &=& \frac{1}{2\pi mk_BT_{\alpha }}.
\end{eqnarray}
With these, the terms for the normalization factor can be re-written as :
\begin{equation}
O(1) ~:~\left(\frac{L}{h} \right)^{2N}2I_{2N}(2mE_{\alpha })
\sum_{m=0}^{\infty } ~\sum_{k=-2m,-2m+2,...2m}\cos (\pi k\nu){\cal P}_{k}^{M}(N);
\end{equation}
and
\begin{eqnarray}
O\left( \frac{h^2}{L^2}\right)~:~\left(\frac{L}{h} \right)^{2(N-1)}
I_{2N}(2mE_{\alpha })B \nonumber \\ 
\sum_{m=0}^{\infty } \sum_{k=-2m-1,-2m+1,...2m+1}~\Bigl[\cos (\pi k\nu){\cal R}_{k}^{M}(N) 
+ \cos (\pi k\nu){\cal Q}_{k}^{M}(N) \Bigr].
\end{eqnarray}
Similarly, the terms corresponding to te numerator of (19) can be re-written as
\begin{equation}
O(1) ~:~\left(\frac{L}{h} \right)^{2N}2I_{2N}(2mE_{\alpha })B\exp (-A)\sum_{m=0}^{\infty } \sum_{k=-2m,-2m+2,...2m}~\cos (\pi k\nu){\cal P}_{k}^{M}(N);
\end{equation}
and
\begin{eqnarray}
O\left( \frac{h^2}{L^2}\right)~:~\left(\frac{L}{h} \right)^{2(N-1)}
&&I_{2N}(2mE_{\alpha })B^2\exp (-A)\nonumber \\
&&\sum_{m=0}^{\infty } \sum_{k=-2m-1,-2m+1,...2m+1}
~\Bigl[ \cos (\pi k\nu){\cal R}_{k}^{M}(N)
+  2\cos (\pi k\nu){\cal Q}_{k}^{M}(N)\Bigr].
\end{eqnarray}

Eqs. (41) and (42) combine to give the numerator of (19) which we call ${\cal F}_1$, 
and, Eqs. (39) and (40) combine to give the denominator of (19) which we call 
${\cal F}_2$. Thus the ratio of ${\cal F}_1$ to ${\cal F}_2 $ gives the momentum distribution upto $O(h^2/L^2)$. Now, after a straightforward arrangement of all the terms , we get
\begin{eqnarray}
\label{F13}
&& F(\vec p_1)=(2 \pi m k T)^{-1} \exp \big( 
-\frac{\vec p_1^2}{2mkT} \big) \bigg\{ 1 \nonumber \\
&& + \big(\frac{h}{L}\big)^2 \frac{1}{2\pi m k T} 
\big( 2 e^{-\frac{\vec p_1^2}{2 m k T}} - 1 \big) G(N, \nu) +
O(\frac{h^4}{L^4}) \bigg\}
\end{eqnarray}
\noindent
where
\begin{equation}
\label{F14}
G(N, \nu) = \frac{\sum_{m=0}^{\infty} \sum_{K=-2m-1 (even)}^{2m+1} 
Q_K^{(m)}(N) \cos(\pi K \nu)   }
{1 + 2 \sum_{m=1}^{\infty} \sum_{K=-2m (odd)}^{2m}
P_K^{(m)}(N) \cos(\pi K \nu)   },
\end{equation}
\noindent
$Q_K^{(m)}$ is the number of elements in $B_N$ composed of $'m'$ generators whereby 
the momentum $\vec p_1$ is interchanged with another momentum yielding a character 
$\exp(i \pi K \nu)$ (or $\exp(-i \pi K \nu)$ since 
$Q_K^{(m)}(N)=Q_{-K}^{(m)}(N)$); $P_K^{(m)}(N)$ is the number of elements in $B_N$ 
contributing to identity with a character $\exp(i \pi K \nu)$ 
( or $\exp(-i \pi K \nu)$). Temperature is introduced above via the ideal gas law,
$E_{\alpha} = NkT_{\alpha}$.
 Unfortunately though, this counting problem stands open today
\cite{17}.  
It is very important to note that the ansatz (\ref{F7})-(\ref{F9}) for the special case when 
$\sigma^2_i=1$ for all $i$ where $B_N$ reduces to symmetric group, $S_N$, the well-known 
Fermi-Dirac and Bose-Einstein distributions follow. On evaluating pressure, $\Pi $ 
from (14), denoting area of the enclosure by A, we get 
${\Pi}A/kT = 1 - (2A)^{-1}{\lambda}^2G(N,{\nu})$, with ${\lambda}^2=h^2(2{\pi}mkT)^{-1}$.
We immediately see that $G(N,0)/(2N)$ and $G(N,1)/(2N)$ are $2^{-3/2}$ and $-2^{-3/2}$
respectively yielding the second virial coeffiecient for the Bose and Fermi gases 
\cite{18}. For the fractional case, with $\nu =$ even number,2j + $\delta $ ("boson-based
anyons"), comparing our result with \cite{8}, we get the Sum Rule :
\begin{equation}
-2^{-3/2}N^{-1}G(N,\nu ){\lambda}^2 = (-1+4|{\delta}|-2{\delta}^2){\lambda}^2/4,
\end{equation}
the right hand side belongs to \cite{8}.
It is important to note that our deduction is non-perturbative and in principle,
 we can get expressions for higher-order virial coefficients also \cite{jm-ko}.
To understand this, we observe that the relation (14) connects two momentum configurations of N particles, and not just the momenta of two particles. Thus, it contains information that can lead to all virial coefficients. For example, for the third virial coefficient, we need to evaluate contributions to $F(\vec p_1)$ when three momenta out of N are interchanged. The denominator of (19) contains those interchanges which braid three strands in such a way that the initial configuration of momenta is preserved whereas the numerator of (19) contains those which exchange the momentum assignment on all three strands. We have done the calculation and the third virial coefficient is expressible in terms of the specific counting problem of $B_N$. Here, in order to convince the reader, it suffices to make a comparative discussion with the existing calculation. For this, we write down the total contribution to the momentum distribution due to a triple interchange emerging from the elements of $B_N$ formed by M generators,
\begin{eqnarray}
\bigg({L \over \hbar}\bigg)^{2(N-2)}& &\sum_{-2M}^{2M}{2 \over 3} I_{2(N-3)}(2mE_{\alpha}-\vec p_1^2)\cos (\pi k\nu ){\cal R}_k^{M}(N) \nonumber \\ &+&2I_{2(N-3)}(2mE_{\alpha}-3\vec p_1^2)\cos (\pi k\nu ){\cal S}_k^{M}(N),
\end{eqnarray}
where ${\cal S}_k^{M}(N)$ (${\cal R}_k^{M}(N)$) are the number of elements of $B_N$ that (do not) change the momentum $\vec p_1$. $I_D(x)$ denotes the volume of a D-dimensional hyper-sphere of radius x. The reason we give this result here is to show that (46) is a Fourier series with harmonic terms like $\cos 2\pi \nu , \cos 4\pi \nu, $ and so on, in complete agreement with the conjectured form \cite{jm-ko}. It is becoming evident from the Monte Carlo calculations \cite{mash} that the third virial coefficient is a series with terms as $\sin ^2\pi \nu , \sin ^4\pi \nu $, and so on. Our formal result is thus in consonance with these works.     
Also, we mention that (45), (46) and the Monte Carlo estimates provide a non-trivial hint on the counting problem
itself.

\noindent
{\bf 4. ~Average entropy of a quantum subsystem - averaging over random eigenstates, and, over random Hamiltonians}

In this section, our discussion will not be restricted to two dimensions. Also, the subject will be entropy which is apparently different from the previous section. The problem that we address is a quantum version of the Ehrenfest urn model \cite{klein} as first considered in \cite{lubkin}.

Consider a system AB with Hilbert space dimension mn and normalized density matrix $\rho $ (a pure state $\rho = |\psi ><\psi |$ if $\rho ^2 = \rho$) \cite{hp,page}. Now we divide this system into two subsystems, A and B, of dimension m and n respectively. The density matrices of A and B, respectively, $\rho _A$ and $\rho _B$, are obtained by partial tracing of $\rho $ over B and A respectively. We assume that A and B are quantally uncorrelated, i. e., $\rho = \rho _A \otimes \rho _B $. If $|\psi >$ is chosen at random, what is the joint probability distribution of eigenvalues of $\rho _A$ ? Following \cite{lubkin}, "random" refers to unitarily invariant Haar measure which, in this case, turns out to be hyper-area of the unit sphere ${\cal S}^{2mn-1}$, the factor 2 coming from the fact that $|\psi >$ has mn complex entries (or 2mn real entries). The objective is to study the average entropy of A, $<S_A>$ (=-tr $\rho _A\log \rho _A$) over the probability distribution of eigenvalues (which are probabilities)  of $\rho _A$. The result of this calculation, conjectured in \cite{page} and proved first in \cite{foong}, is
\begin{equation}
<S_A> = \sum_{k=n+1}^{mn} {1 \over k} - \frac{m-1}{2n},~~~~~~~~(m \leq n).
\end{equation}

The random pure state can be written (for the system we are considering) as (7,8) in three (or greater) space dimensions, or, as (13-15) in two dimensions. These equations can be interpreted as choosing a pure state at random for a specific choice of amplitude, $A_{\alpha }$. 

If one calculates the average of trace of $\rho _A^2$, first, over homogeneously distributed unit vector in mn-dimensional Hilbert space, and then, over random Hamiltonians, the two answers are only different by one bit \cite{eli1}. The values are almost the same as one corresponding to the answer when the entropy will be maximal, i.e., $\log m$. This brings about the random matrices as all that is being done here about averaging over random Hamiltonians is what is done in random matrix theory. Thus the average entropy of a subsystem follows from the random matrix hypothesis about AB - the statement becomes exact when $m \ll n$. Indeed, the connection of random matrix theory and statistical mechanics is when the size of the system is large where it means then that the number of particles is large to be consistent with thermodynamic limit. 

It is very interesting to note that if the pure states of AB are random, the probability distribution of eigenvalues of $\rho _A$ is just the one-point correlation function corresponding to the (generalized) Laguerre unitary ensemble of random matrices \cite{ruiz}. Since the one-point correlation function (average level density) is the same for orthogonal, unitary, and symplectic ensembles \cite{slevin}, the answer for the average entropy will remain the same. Let us remember that the average level density is, in principle, a function of the size of the matrices. The fact that we must discuss entropy in the context of systems in thermodynamic limit is what makes the entropy same for all the ensembles. Since the fluctuations on top of the average density become significant with decreasing sizes, we expect to observe their interesting effect on the entropy. We now present results that prove these remarks.

We begin by recalling that the average over all pure states of AB, in unitary Haar measure, of the spread of eigenvalues of $\rho _A$ is \cite{lubkin}
\begin{eqnarray}
<\sigma ^2> &=& <{1 \over m}\sum_{i=1}^{m} \left( p_i - \frac{1}{m} \right)^2> \nonumber \\
&=& \frac{1-m^{-2}}{mn+1}.
\end{eqnarray}
Clearly, the case n=1 corresponds to the situation when $\rho _A$ is also pure, then we have 
\begin{equation}
<\sigma _{max}^2> = \frac{1-m^{-2}}{m+1},
\end{equation}
the ratio of $<\sigma ^2>$ to $<\sigma _{max}^2>$ gives the measure of "purity" of the subsystem A. Since we have noted above that the same answer can be obtained by averaging over the one-point correlation function of the Laguerre unitary ensemble of random matrices, we expect that apart from the leading term which is $\log m$ for the entropy as this corresponds to equipartition, the "defect" term must show the signature of different ensembles. To this end, we start by writing the entropy,
\begin{equation}
S = - \sum_{i} p_i \log p_i,
\end{equation}
and Taylor expand each $p_i$ about $1/m$. With 
\begin{equation}
p_i = \frac{1}{m} - \frac{q_i}{m},
\end{equation}
we can write
\begin{equation}
S = \log m - \frac{1}{m}\left[ \frac{1}{1.2}\sum_{i=1}^{m}q_i^2 + \frac{1}{2.3}\sum_{i=1}^{m} q_i^3 + ... \right],
\end{equation}
which is convergent if $|q_i| < 1$, i.e., if $0 < p_i < \frac{2}{m}$. Since $<\sigma ^2>$ is small for large n, most of the measure will lie with $p_i < \frac{2}{m}$. It is plausible that $S = \log m -\mbox{defect}$, and that the defect is well approximated by 
\begin{eqnarray}
<\mbox{defect}> &\equiv & \frac{1}{2m} \sum_{i} q_i^2 = {1 \over 2} m^2\sigma ^2 \nonumber \\
&=& {1 \over 2}\frac{m^2 - 1}{mn + 1}.
\end{eqnarray}
We can now find the defect for the case where we integrate over orthogonally invariant Haar measure \cite{canada} and symplectically invariant Haar measure. The difference is that these correspond to ${\cal S}^{mn-1}$ and ${\cal S}^{4mn-1}$ respectively. After the same steps, we get
\begin{equation}
\left< \frac{\sigma ^2 }{\sigma _{max}^2}\right> = \frac{{\beta \over 2}m+1}{{\beta \over 2}mn+1}
\end{equation}
where $\beta $ is the co-dimension of level crossing, respectively 1,2, and 4 for orthogonal, unitary, and symplectic ensembles of random matrix theory. The entropy is given by 
\begin{equation}
S \equiv \log m - \frac{(m-1)}{2}\frac{{\beta \over 2}m+1}{{\beta \over 2}mn+1}.
\end{equation}
This result shows that the entropy is almost maximal for n large enough, and that the finite-dimensional effects show up the dependence on the global symmetries of the system. 

We wish to note that the arguments we have used are quantum mechanical. The results of this section show that randomness in eigenstate, which follows from random matrix theory, encompassing the ergodicity of the classical system leads to entropy of a subsystem  which is maximal. To notice the differences between the chaotic quantum systems belonging to different random matrix ensembles, we need to study the entropy as a function of the Hilbert space dimension as is clear above.  We conclude this section with the mention of a recent work where a related study is  carried out on periodically kicked top \cite{amm}.

\noindent
{\bf 5.~ Concluding  Remarks}

We have shown in this paper that an ansatz where eigenstates are written as random 
superpositions of plane waves (or some basis consistent with the boundary conditions) 
for systems whose classical analogues are chaotic is equivalent to a random matrix 
hypothesis. In an important work \cite{4}, the connection between the ansatz and 
statistical mechanics in three dimensions was brought out. Since the group that governs 
exchange symmetry in two dimensions is an infinite,  nonabelian  one whose special case 
is the permutation group (the exchange group in three dimensions), our treatment and 
results in Section 3 are  generalizations of  \cite{4}. It was possible for us to do 
this only because we realised what the random matrix ensemble should be in two dimensions. 
This is the reason for Section 2 where a comparative discussion about the relevance of 
space dimensions is given to the random matrix ensembles. In consistency with the 
expectations, the ensemble in two dimensions is chiral unitary ensemble. It is well-known 
that choosing a specific nature of randomness (e.g., Gaussian) gives then the 
average density of states which is not realistic. This can be treated with the Dyson 
Brownian motion model \cite{12}  where any realistic density of states can be modelled. 
An interesting relation between a generalized Brownian motion model and a semiclassical 
reasoning of universality has been recently worked out \cite{me} by a generalization of the theory of level dynamics. 

The fact that both the ideas, one  of random pure state of an isolated system, and, 
that of this system being governed by a random Hamiltonian, give rise to the 
distribution functions and virial coefficients correctly suggest that there may be a 
connection between the two. In Section 4, we have described a way that we see most 
clearly (as of now) in the context of entropy of a quantum subsystem. We believe that 
the arguments developed here provide a common ground to seemingly different themes, leading  to the second law of thermodynamics. 

As shown in Section 3, the number of words formed by M generators of the Braid group 
satisfy a sum rule which comes from our calculation of the second virial coefficient. 
The parameter, $\nu $ has an analogous 
partner in quantum chromodynamics \cite{8} and we conjecture that the anyon gas 
discussed here and the $\nu \pi $-parametrised quantum chromodynamics belong to 
the same universality class of chiral Unitary Ensemble of RMT.
 
\newpage
\noindent
{\bf Acknowledgements}
 
S.R.J. is financially
supported by the "Communaute Francaise de Belgique" under contract no.
ARC-93/98-166. S. R. J. thanks Elihu Lubkin for stimulating electronic discussions on several aspects on entropy; he also thanks Mark Srednicki for a number of discussions and for a very hospitable stay at the University of California, Santa Barbara.

\end{document}